\documentclass[prb,twocolumn,showpacs,showkeys,preprintnumbers,superscriptaddress,amsmath,amssymb]{revtex4}% P.R.L.

\usepackage{color}%
\usepackage{graphicx}% Include figure files
\usepackage{bm}% bold math
\usepackage{wasysym}

\newcommand{\lamno}{LaMn$_7$O$_{12}$}
\newcommand{\lamnoS}{LaMnO$_{3}$}
\newcommand{\lamnoO}{LaMnO$_{3.02}$}

\newcommand{\lacamno}{La$_{0.7}$Ca$_{0.3}$MnO$_3$}

\begin{document}

\title{The high temperature Jahn-Teller transition in \lamno}

\author{R.~Cabassi}
\author{F.~Bolzoni}
\author{E.~Gilioli}
\author{F.~Bissoli}
\affiliation{Istituto dei Materiali per Elettronica e Magnetismo IMEM-CNR, 
Parco Area delle Scienze 37/A,I-43100 Parma, Italy}
%\email[Corresponding Author.~E-mail: ]{<cabassi@imem.cnr.it>}

\author{A.~Prodi}
\affiliation{Istituto dei Materiali per Elettronica e Magnetismo IMEM-CNR, 
Parco Area delle Scienze 37/A,I-43100 Parma, Italy}
%\affiliation{Department of Physics, Massachusetts Institute of Technology, Cambridge, MA 02139-4307, USA.}

\author{A.~Gauzzi}
\affiliation{Istituto dei Materiali per Elettronica e Magnetismo IMEM-CNR, 
Parco Area delle Scienze 37/A,I-43100 Parma, Italy}
\affiliation{Institut de Min\'eralogie et de Physique des Milieux Condens\'es,
Universit\'e Pierre et Marie Curie, CNRS 140, rue de Lourmel, 75015 Paris France}

% \author{Q.~Huang}
% \author{A.~Santoro}
% \author{J.W.~Lynn}
% \affiliation{NIST Center for Neutron Research, Gaithersburg, Maryland 20899, USA}
% 
% \author{M.~Affronte}
% \affiliation{CNR-INFM-S3 and Dipartimento di Fisica, Universit\`{a} di Modena e Reggio Emilia, Modena 41100, Italy}

% \author{E.~Gilioli}
% \author{F.~Licci}
% \affiliation{Istituto dei Materiali per Elettronica e Magnetismo IMEM-CNR, 
% Parco Area delle Scienze 37/A,I-43100 Parma, Italy}
% \author{M.~Marezio}
% \affiliation{Istituto dei Materiali per Elettronica e Magnetismo IMEM-CNR, 
% Parco Area delle Scienze 37/A,I-43100 Parma, Italy}
% \affiliation{CRETA CNRS, 38042 Grenoble cedex 9, France}
% \author{F.~Bolzoni}
% \author{R.~Cabassi}
% \affiliation{Istituto dei Materiali per Elettronica e Magnetismo IMEM-CNR, 
% Parco Area delle Scienze 37/A,I-43100 Parma, Italy}
% \author{A.~Gauzzi}
% \affiliation{Istituto dei Materiali per Elettronica e Magnetismo IMEM-CNR, 
% Parco Area delle Scienze 37/A,I-43100 Parma, Italy}
% \affiliation{Institut de Min\'eralogie et de Physique des Milieux Condens\'es,
% Universit\'e Pierre et Marie Curie, CNRS 140, rue de Lourmel, 75015 Paris France}

\date{\today}

\begin{abstract}

A first order structural Jahn-Teller transition at $T_{JT} \approx 650 K$ has been recently reported for the quadruple perovskite \lamno\ .
We have carried out magnetization and transport measurements below and above  $T_{JT}$ in order to investigate the effect of the transition.
Electrical conduction turns out to be polaronic, changing from non-adiabatic to adiabatic through the transition.
 Magnetic behavior can be described by non-interacting Mn$^{3+}$ ions below $T_{JT}$, while above $T_{JT}$ it is of questionable interpretation.
The effect of thermal cycling on as grown samples of different purity degree also allowed us to clarify the intrinsic magnetic response of \lamno\ at lower temperatures.

\end{abstract}


% 75.30-m     Intrinsic properties of magnetically ordered materials
% 75.30.Vn    Colossal magnetoresistance
% superstructure

% 61.50.Ks    Crystallographic aspects of phase transformations;
% pressure effects
% 61.66Fn     Structure of specific crystalline solids: Inorganic Compounds

% 61.10.Nz X-ray diffraction
% 61.12.Ld neutron diffraction
% 61.14.-x electron diffraction and scattering

\maketitle

\section{Introduction \label{intro}}

Quadruple perovskites of $AA'_3B_4$O$_{12}$ structure~\cite{Mare,Boch} are derived from the conventional $AB$O$_3$ perovskite structure by means of the chemical cell axes doubling. When both the $A'$ and $B$ site are occupied by Mn the family $AMn_7$O$_{12}$ is obtained. In this family, a wide variety of interesting properties involving charge, spin and orbital ordering is observed, depending on the ion occupying the $A$ site: some examples are given by~\cite{NAMNO,CAMNO,PRMNO,BIMNO,LAMNO}   $A$=Na,Ca,Pr,Bi and La. 
The single valent Mn oxide \lamno\ in particular is of special interest because of the relevance of its simple perovskite analogue \lamnoS\ , which is the parent compound of the well studied family of the colossal magnetoresistance (CMR) doped manganites.
The properties of \lamno\ have already been studied in detail  in the temperature range below room temperature~\cite{LAMNO} (RT), with special regard to its peculiar magnetic ordering consisting of two independent sublattices, namely the $B$ sites order according to a C-type antiferromagnetic pattern and the $A'$ sites order with antiferromagnetically coupled ferromagnetic planes.

Both \lamno\ and  \lamnoS\ are characterized by similar pseudocubic networks of buckled corner sharing MnO$_6$ octahedra subjected to cooperative Jahn-Teller distorsion. However, in \lamno\ the small Mn-O-Mn bond angle gives rise to very different magnetic properties below RT.
Above RT, the MnO$_6$ octahedra in \lamnoS\ loose their Jahn-Teller distorsion at 750 K where an orthorhombic to rhombohedral structural transition takes place~\cite{Rodr97,Rodr}, the effect of this transition over magnetic and transport properties has been the subject of several authors~\cite{Zhou2,Mandal,Souza1,Zhou3}. A similar structural transition from monoclinic ($I2/m$) to cubic ($Im3$)    has been recently observed~\cite{Okam} in \lamno\ by means of synchrotron X-ray diffraction experiments at  $T_{JT} \simeq 650 K$ , where the Jahn-Teller distorsion of the MnO$_6$ octahedra is relaxed.

The aim of this paper is to study the magnetic and transport properties of \lamno\ above RT, analyzing how do they change while crossing through the Jahn-Teller transition at $T_{JT}$ . The performed magnetic measurements also offer the opportunity to clarify an open question left by the low temperature study~\cite{LAMNO} regarding  anomalies observed in magnetization and in structural parameters at $T \simeq 200 K$ .

\section{Experimental Methods \label{exp}}

% \subsection{Synthesis}
The synthesis of \lamno\ was carried out at high pressure, the details have been published elsewhere~\cite{LAMNO}.
Magnetic properties were characterized using a Quantum Design SQUID magnetometer equipped with 5.5T magnet and high-temperature oven. The electrical resistivity was measured in the four-probe (van der Pauw) configuration.

\section{Results \label{res}}

\subsection{Magnetic Properties}

\subsubsection{Effect of heat treatment and secondary phases on magnetism}

In the analysis of magnetic measurements of \lamno\ samples, special care should be taken considering the effect of impurities that can be easily result from the synthesis. In order to attain insight in this aspect, different polycristalline samples of LaMn$_7$O$_{12}$ have been characterized, high quality samples grown in optimal conditions and low quality ones grown in non perfect conditions of pressure and temperature. Here we present the results of two samples: an high-purity sample (in the following referred to as $S1$) and a low-purity sample (in the following referred to as $S2$). The open question of the recent paper ~\cite{LAMNO} regarding the presence of a secondary phase guessed as \lamnoS\ is analyzed more accurately in the following. 

\begin{figure}
	\centering
	\includegraphics[width=80mm]{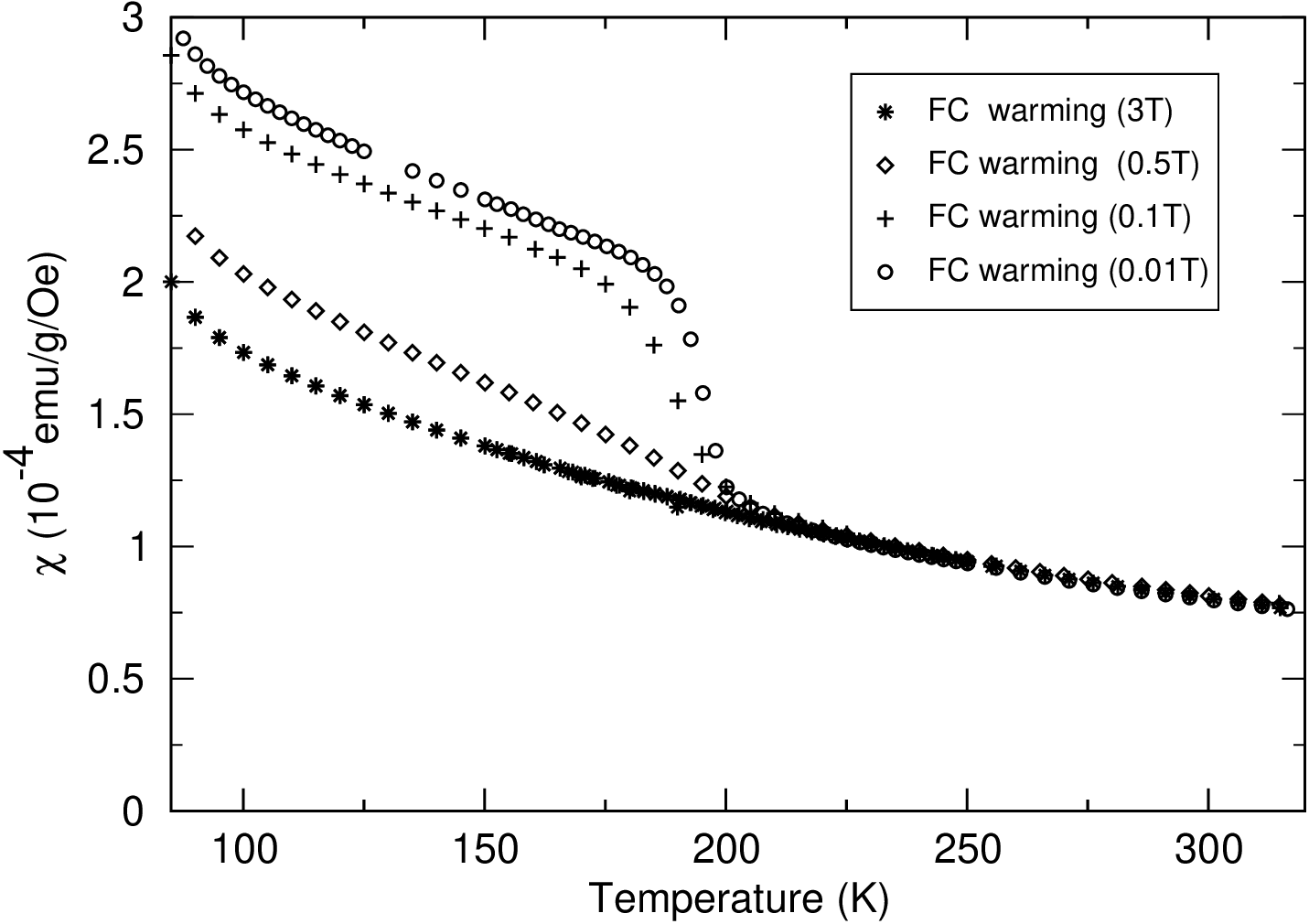}
	\caption{Magnetic susceptibility of LaMn$_7$O$_{12}$ for different values of applied magnetic field.}
	\label{fig:lamnospurio2}
\end{figure}

Fig. \ref{fig:lamnospurio2} shows the magnetic susceptibility of the sample $S1$ obtained in field cooling/warming (FCW) mode for various applied fields. It is interesting to note that as the signal from the guessed \lamnoS\ secondary phase saturates with increasing field, its percentage contribution  becomes smaller and the total susceptibility tends to the Curie-Weiss law of the main phase. 

\begin{figure}
	\centering
	\includegraphics[width=80mm]{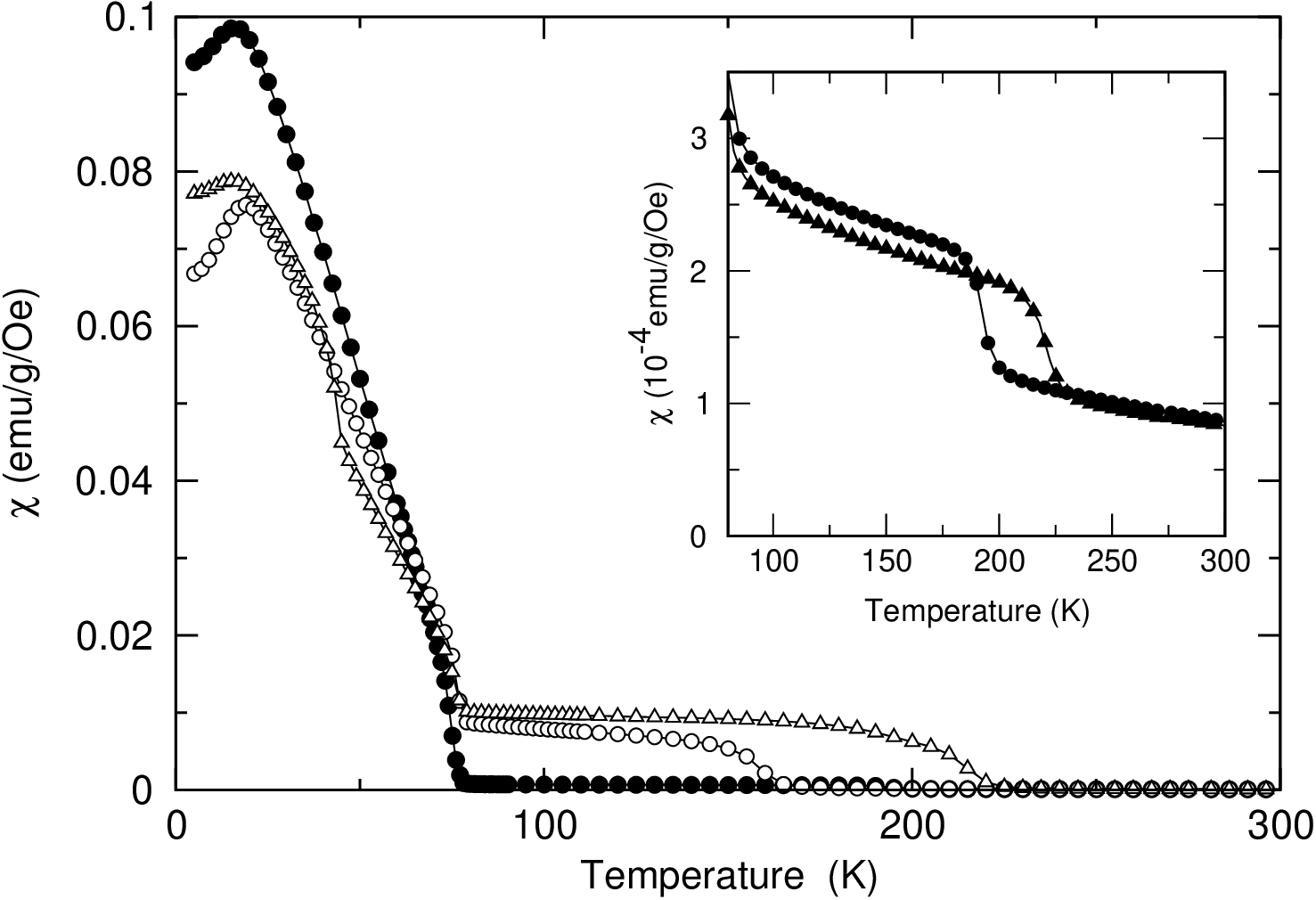}
	\caption{Magnetic susceptibility of a high ($S1$) and a low ($S2$) quality \lamno\ sample  obtained in 100 Oe field cooling warming (FCW) mode. $\CIRCLE$:\ $S_1$ as grown; $\Circle$:\ $S_2$ as grown; $\triangle$:\ $S_2$ treated at 1173 K. Inset: expanded scale for the high quality sample $S_1$ in the transition region of the secondary phase. $\CIRCLE$:\ $S_1$ as grown; $\blacktriangle$:\ $S_1$ treated at 800 K.}
	\label{fig:lamnospurio1}
\end{figure}

Comparison of the field warming magnetization in 100 Oe of the high purity sample $S1$ (full symbols) and the low purity sample $S2$  (empty symbols) is shown in Fig. \ref{fig:lamnospurio1}. S2 sample displays, in addition to the known transitions at low temperature corresponding to the magnetic ordering of B and A' sublattices, a third transition at T $\approx$ 200 K, while in the S1 sample the guessed secondary phase is not detected on this scale.
Inspection of X-ray diffractograms on sample $S2$ allows to associate this signal to a secondary phase with \lamnoS\ structure.

\begin{figure}
	\centering
	\includegraphics[width=80mm]{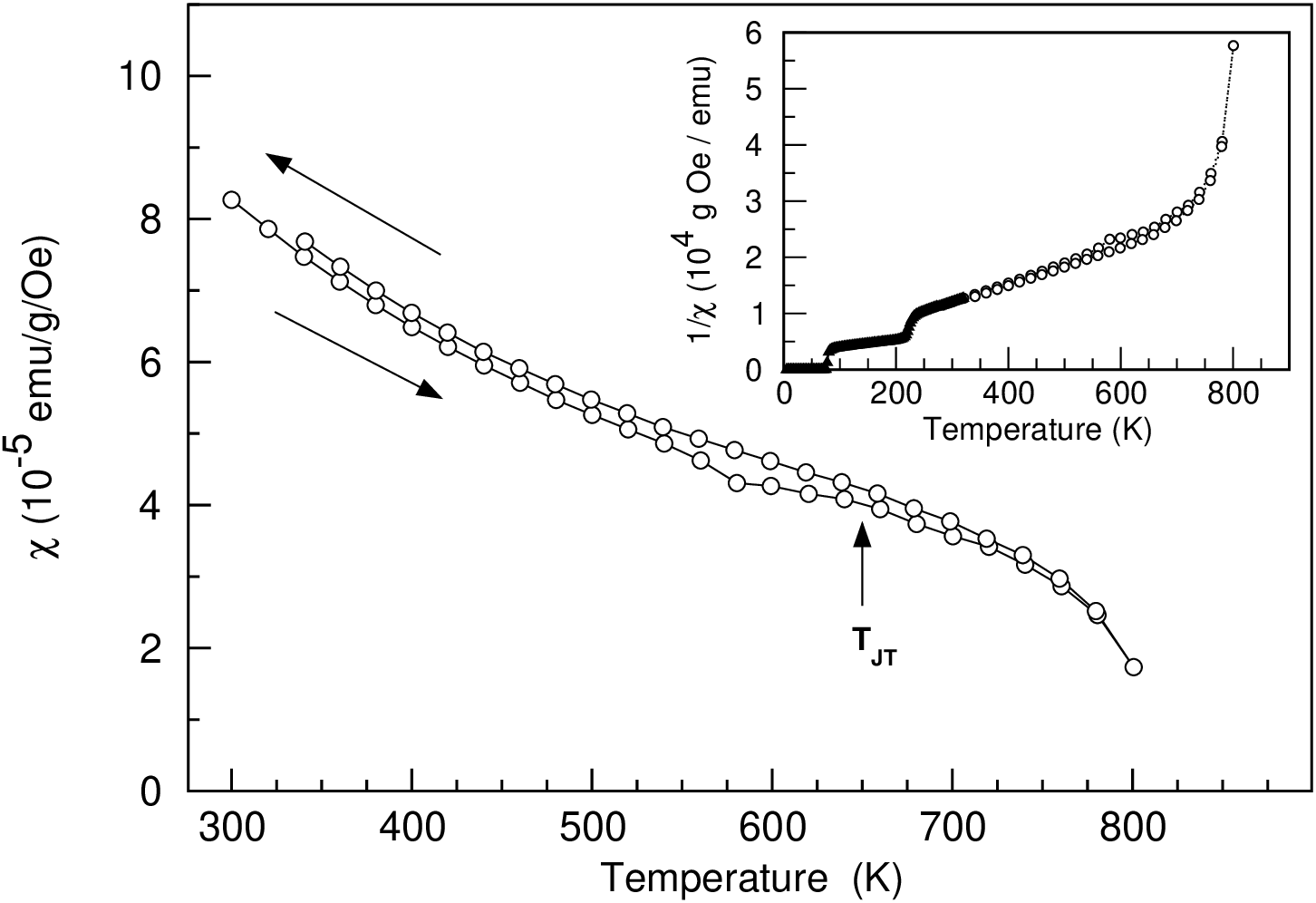}
	\caption{Magnetic susceptibility of LaMn$_7$O$_{12}$ as a function of temperature measured in a field of 100 Oe. Inset: inverse susceptibility in the whole temperature range, as resulting from two separate measurement sessions below and above Room Temperature respectively. The shape between 78 K and 200 K reflects the presence of \lamnoS\ impurity.  }
	\label{fig:lamnomt}
\end{figure}

The transition temperature of the secondary phase in $S2$ changes on heat treatment, namely we measured $T_c$ = 160 K for as grown $S_2$ and  $T_c$ = 215 K for $S2$  heat treated at 1173 K, while the other transitions of \lamno\ remain unaffected. Variations of \lamnoS\  transition temperature are well known in literature and are associated both to oxygen off-stoichiometry~\cite{Topf, Topf2, Tiwa, Muro, Vere} and to pressure~\cite{Zhou1}, being stoichiometric \lamnoS\  antiferromagnetic with Ne\'el temperature $T_N = 140 $ K. Relaxation from after-effects of the hydrostatic pressure acting during the synthesis procedure would decrease the transition temperature~\cite{Zhou1}, therefore we can likely ascribe the shift in the transition temperature of $S2$ to changes in stoichiometricity induced by heating, nonetheless it has been shown~\cite{Joy} that it is not possible to estabilish a definite correlation between the transition temperature and the amount of oxygen excess.
The slope of inverse susceptibility in the paramagnetic range (inset of Fig. \ref{fig:lamnomt}) yield 4.98 and 4.93 $\mu_B$ / Mn ion for as grown and heat treated $S1$ respectively, with Weiss constant $\theta$ = -48 K and $\theta$ = -54 K. Conversely, for $S2$ the number of Bohr magnetons is lower and  increases on temperature treating, suggesting a decrease in Mn$^{4+}$ ions in the impurity phase and pointing again to stoichiometricity variations.
The inset in Fig. \ref{fig:lamnospurio1} shows on an expanded scale the magnetization of sample $S1$ in the transition region of the secondary phase. It reveals that a much weaker contribution to the magnetization from the secondary phase is still present even when, as in the present case, 
it cannot be detected neither from X-ray nor neutron diffraction, and its relative fraction can be estimated to be below 1 $\%$. In this case as well, a heat treatment at T = 800 K results in a shift to lower temperatures of the magnetic transition.

The above outlined analysis allows us identify the origin of the susceptibility anomaly at T $\approx$ 200 K with the presence of \lamnoS\  as secondary phase, with different amount in the measured samples. 
It is also manifest that the low temperature features of susceptibility are not affected by heat treatment, and that therefore in \lamno\ the magnetic response is quite more reproducible than in the parent compound \lamnoS\ . One can also remark that the presence of \lamnoS\  impurity in \lamno\ samples explains the structural anomaly at $T = 200$ K already observed~\cite{LAMNO} in the lattice parameters as function of temperature.

\subsubsection{High-temperature transition}

The magnetic susceptibility of sample $S1$ measured in a field of 100 Oe in the high temperature region is displayed in Fig.~\ref{fig:lamnomt}, together with its inverse. The inverse susceptibility shows that the Curie-Weiss law is precisely followed up to $T \le$ 600 K, yielding 4.94 and 4.99 $\mu_B$/Mn ion for warming and cooling measurement respectively, with Weiss constant $\theta$ = -57 K and -53 K. 
These values are almost equal to the results of  measurements performed without the high-temperature oven displayed in Fig.~\ref{fig:lamnospurio2} supporting therefore the reliability and reproducibility of results, as is also clear from inset of Fig.~\ref{fig:lamnomt} where the inverse susceptibilities of the two measurements are seen to match each other perfectly.
 The negative Weiss constant reflects the prevalence of antiferromagnetic with respect to ferromagnetic interactions as an overall result of the particular magnetic ordering of the two sublattices~\cite{LAMNO} $A'$ and $B$.
Two major features are noteworthy in Fig.~\ref{fig:lamnomt}, the first one is a small thermal hysteresis at $T \approx$ 600 K,the second one is the abrupt fall in magnetic signal towards T = 800 K. These features are close to the Jahn-Teller transition, and are discussed in Section~\ref{sec:disc}.

\begin{figure}
	\centering
	\includegraphics[width=80mm]{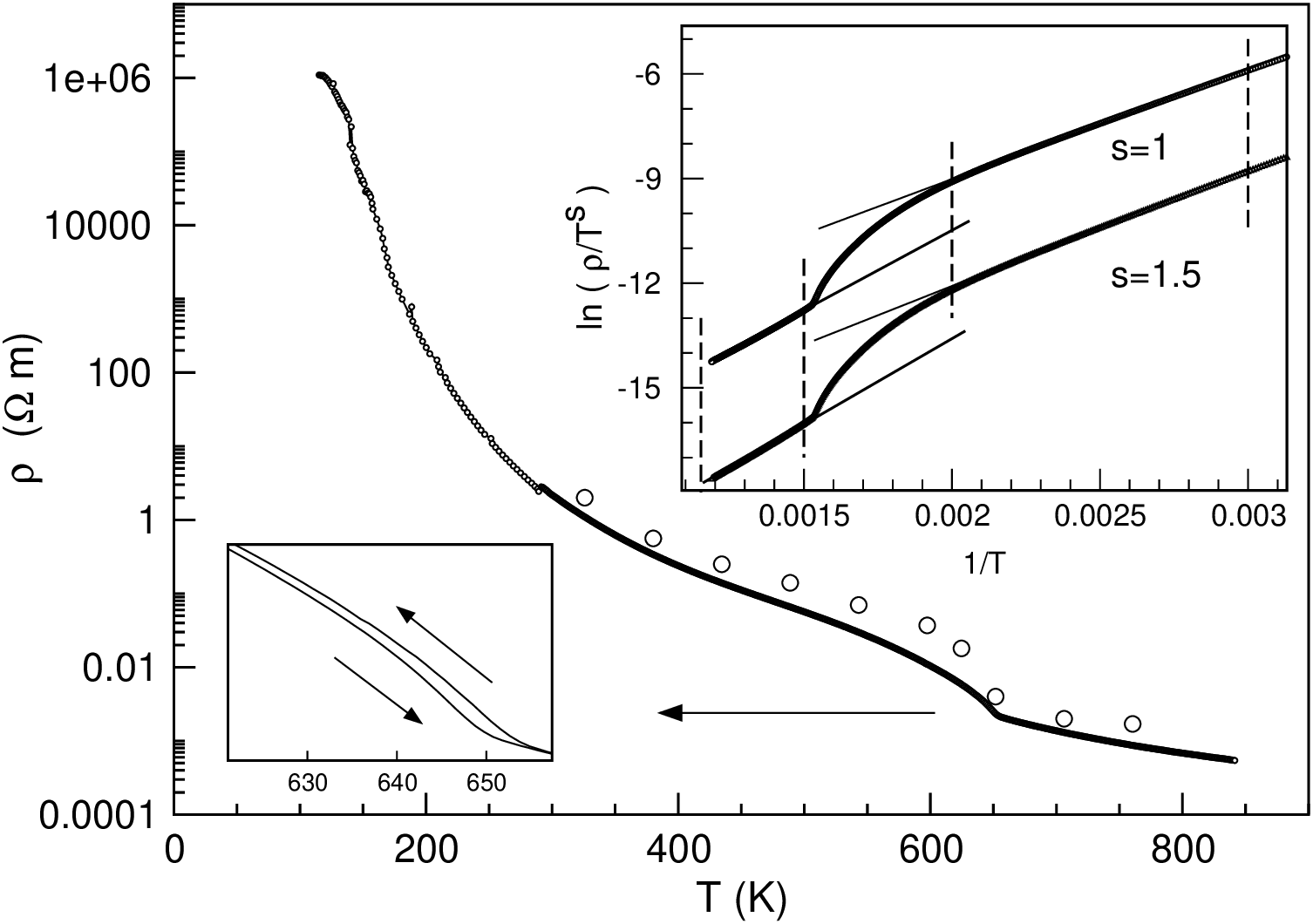}
	\caption{Temperature  dependence of the dc electrical resistivity of \lamno\ .  $\Circle$: selected resistivity values of \lamnoO\  extracted from Ref.~\cite{Souza1} and scaled for the different values of the structural transition temperatures of the two compounds. Upper inset: $1/T$ plots for adiabatic ($s$=1) and non-adiabatic  ($s$=1.5) polaron conduction in the range $T > \Theta_D/2 $. Solid lines are linear best-fits, dashed lines  define range for fits. Lower inset: hysteretic behavior of the transition.}
\label{fig:lamno_rho}
\end{figure}

\subsection{Transport Properties}
The dc electrical resistivity is reported in Fig.~\ref{fig:lamno_rho}, where the transition at $T_{JT}$ is clearly visible with a small hysteresis enlarged in lower inset. 
It is worth to remark that results are reproducible after temperature cycling up to 850 K, so that they cannot be ascribed to variations in the oxidation state of the sample.

We propose to model the measured data on the basis of lattice small polarons that can be formed by the $e_g$ electrons of the MnO$_6$ octahedra in the sublattice $B$. This is the same mechanism that has been found to be at work in \lamnoS\ , as well as in members of the realated family of doped CMR manganites.
\begin{table*}
\caption{\label{tab:tableFit}Parameters for adiabatic and non-adiabatic small polaron fit to electrical conductivity of \lamno\ .}
\begin{ruledtabular}
\begin{tabular}{llclc}
 & &$E_{\sigma}(meV)$&$\sigma_0 T_0^s(\Omega^{-1}m^{-1}K^s)$&Range for fit (K)\\
\hline
$T<T_{JT}$ &\ &\ &\  & [330;500]\\\
 \ &Non-adiabatic limit & 290 & 1.5\ 10$^8$ &\\
 \ &Adiabatic limit & 270& 4.5\ 10$^6$ &\\
$T>T_{JT}$ &\ &\ &\ & [670;840]\\
 \ &Non-adiabatic limit & 430 & 1.8\ 10$^{10}$ & \\
 \ &Adiabatic limit & 400& 4.0\ 10$^{8}$ & \\
\end{tabular}
\end{ruledtabular}
\end{table*}
In order to carry out the analysis of the results, we remind that polaron hopping conductivity is thermally activated according to the expression~\cite{emin, austin}
\begin{equation}
\sigma = \sigma_0 \left(\frac{T_0}{T}\right)^s exp\left(- \frac{E_{\sigma}}{k_BT}\right)
\label{eq:sigma}
\end{equation}
holding for  $T > \Theta_D / 2$ where $\Theta_D$ is the Debye temperature. The exponent can be $s=1$ for the adiabatic limit, when lattice distorsions are slower than hopping frequency and all hopping attempts are successful, or  $s=1.5$ for the non-adiabatic limit, when lattice distorsions are faster than hopping attempts.
 Inside inset of Fig.~\ref{fig:lamno_rho} are shown the $1/T$ plots for $ln(\rho/T)$ and $ln(\rho/T^{3/2})$ in the range $T > \Theta_D/2 \simeq 330 K$, where $\Theta_D$ is the Debye temperature~\cite{LAMNO} of \lamno\ . Linear best-fits were performed in the range for fit outlined by dashed lines drawn in Fig.~\ref{fig:lamno_rho}, and are in very good agreement with experimental data for both adiabatic and  non-adiabatic limit, below and above $T_{JT}$, therefore one cannot choose the correct hopping regime on the basis of the quality of best-fits.The fit parameters are reported in Table~\ref{tab:tableFit}. It is however possible to gain deeper insight~\cite{Jaime} recalling that in the adiabatic limit
\begin{equation}
%\begin{multline}
%\begin{aligned}
\sigma_0 = \ \frac{g_{d} e^2}{a \hslash}\\
\label{eq:adiabLimit1}
%\end{aligned}
%\end{multline}
\end{equation}
and
\begin{equation}
%\begin{multline}
%\begin{aligned}
k_B T_0 = \ h \nu_0
\label{eq:adiabLimit2}
%\end{aligned}
%\end{multline}
\end{equation}
where $a$ is the hopping distance, $g_d$ is a factor depending on the hopping geometry and $\nu_0$ is the characteristic phonon frequency, thus at temperature $T \simeq T_0$  the prefactor $\sigma_0 T_0 / T$ can be directly obtained by eq.~\ref{eq:adiabLimit1}. If we take for the hopping distance $a$ the Mn-Mn length of 0.32 nm resulting from structural data reported in Ref.~\cite{LAMNO}, we obtain $\sigma_0 T_0 / T \simeq 1.2 \ 10^5 g_d (\Omega \ m)^{-1}$. A prefactor approaching this value is the signature of adiabatic hopping regime, while a lower value would correspond to non-adiabatic regime. Considering that the possible range for $g_d$ is~\cite{Jaime} from $g_d=1$ to $g_d=5$ (for nearest neighbour on square planar lattice and next-nearest neighbour hopping respectively), an inspection of Table~\ref{tab:tableFit} suggests that hopping processes are adiabatic for $T > T_{JT}$, while for $T < T_{JT}$ hopping is non-adiabatic. Hopping activation energy is 0.29 eV below $T_{JT}$, while it amounts up to 0.40 eV above $T_{JT}$.

\section{Discussion \label{sec:disc}}
The  monoclinic to cubic structural transition of \lamno\ at $T_{JT} \approx$ 650 K is very clearly visible in both transport and magnetic measurements, which also show a corresponding thermal hysteresis confirming the first order nature of the transition. The discussion of results can benefit from a comparison with the analagous orthorhombic to rhombohedral transition of the parent compound \lamnoS\ , which has been analyzed by some authors for pure single crystals~\cite{Zhou2}, almost pure \lamnoO\  polycrystalline samples~\cite{Souza1} and doped polycrystalline~\cite{Souza2} \lacamno\ . 

Electrical resistivity of \lamno\ is semiconducting-like ($d\rho / dT < 0$) and polaronic in nature both below and above $T_{JT}$, with a crossover from non-adiabatic to adiabatic hopping regime when passing to higher crystal symmetry. This behavior is very similar to what has been found for lightly doped~\cite{Souza1,Mandal} or off-stoichiometric~\cite{Souza2} samples of \lamnoS\ (while on the other hand pure \lamnoS\  single crystals pass from polaronic to temperature independent resistivity, originated from itinerant vibronic charge carriers having fixed mean free path~\cite{Zhou2, Zhou3}), and could be explained by the presence of a small amount of  Mn$^{4+}$ present in our sample.
Jahn-Teller fluctuations above $T_{JT}$ can also introduce more Mn$^{4+}$ ions by means of charge disproportionation 2Mn$^{3+}$ = Mn$^{2+}$ + Mn$^{4+}$ on some of the Mn atoms. 

The hopping energy we have measured for \lamno\ at $T < T_{JT}$ is practically equal to the one measured~\cite{Souza1} for \lamnoO\ below and above $T_{JT}$, which amounts to 0.29 eV and 0.28 eV respectively.
This strict similarity suggests that the polaronic transport is supported by the sublattice $B$ only, which makes sense according to the more covalent character of the Mn-O bonds of the $A'$ sublattice~\cite{LAMNO}. To strengthen this point we have reported in Fig.~\ref{fig:lamno_rho} some selected values of the electrical resisitivity (on cooling) of \lamnoO\  extracted from Ref.~\cite{Souza1} and scaled by the ratio of the structural transition temperatures of two compounds: the values are very similar, whilst if the $A'$ sublattice of \lamno\ took part to the formation of lattice polarons, one would expect a variation in polaronic mass hence a change in resistivity.
%This strict similarity confirms that transport is supported by lattice polarons formed by $e_g$ electrons of the $B$ sublattice, and suggests that the energy band structure of the $A'$ and `$B$ sublattice does not affect each other in the monoclinic phase.
Magnetization measurements give an inverse susceptibility strictly following a linear Curie-Weiss law in the temperature range from the magnetic transition of the \lamnoS\ impurity up to $T_{JT}$. The resulting paramagnetic moment per Mn ion is equal within 1 $\%$ to the value of 4.89 $\mu_B$ expected for non-interacting Mn$^{3+}$ ions (or 5.00 $\mu_B$ in the case of Mn$^{2+}$Mn$^{4+}$ disproportionation pairs). Therefore it looks likely that the $A'$ sublattice does not alter the charachter of behavior below the transition, and that its sole role is to lower the temperature at which the transition to the higher strutctural phase sets in.

Conversely, in the cubic phase we measured a greater hopping energy amounting to 0.40 eV; similar enhancement in the activation energies above $T_{JT}$ has already been observed in some single-valent manganites~\cite{Zhou3}, and has been explained by the energy required to create disproportionation fluctuations.
The magnetization in the cubic phase exhibits a peculiar drop, difficult to discuss because it takes place near the upper bound of the range accessible by the instrumentation. Ferromagnetic impurity  with Curie temperature close to 800K can be ruled out because of the perfect linearity of $\chi^{-1}$ in the temperature range from T$_N$ to $T_{JT}$, and equally the thermal hysteresis size is too small to ascribe the drop in magnetization to variations in the off-stoichiometry oxygen content. The apparition of unquenched orbital moment in cubic symmetry, even though unusual, cannot be completely ruled out, as recently shown for localized impurities in bulk metal with cubic symmetry \cite{Brever:04} and for strained \lacamno\ films \cite{Song:05}, however in order to validate this hypotesis by means of magnetization measurements one should  heat the sample until the linear Curie-Weiss regime is recovered, well further the accessible range. On the other hand, synchrotron X-ray diffraction results~\cite{Okam} rule out the possibility of further structural transitions above $T_{JT}$, but also point to a transition from static Jahn-Teller distorsions to dynamical fluctuations around average values with cubic symmetry, in a similar way to what observed~\cite{Rodr97} in \lamnoS\ . We argue that this effect, together with the possible coexistence of volume fractions of the two phases in some temperature range around $T_{JT}$, could be at the origin of the observed feature as well.

\section{Conclusions \label{concl}}
We have measured magnetic and transport properties of \lamno\ in the temperature range up to 800 K, observing in this way the effect of the first order structural Jahn-Teller transition at  $T_{JT}$ = 650 K. Electrical conductivity is supported by small polaron hopping, with a crossover from non-adiabatic to adiabatic regime on heating through $T_{JT}$. Magnetic measurements  below  $T_{JT}$ show the ordinary behavior of paramagnetic non-interacting ions, while above  $T_{JT}$ it is observed a puzzling deviation from Curie-Weiss regime, which can be associated to the change of the bonding angles, hence of the exchange integrals, at $T_{JT}$. By means of magnetic analysis of different samples, we also have clarified the role of \lamnoS\ impurities below RT. In summary, transport properties look very similar in \lamno\ and in the simple perovskite analogue \lamnoS\ , while the magnetic properties of \lamno\ show a peculiar behavior above  $T_{JT}$, which is connected to the change of symmetry properties and would need measurements at higher temperatures to be better understood.

\bibliography{manga}

\end{document}